\documentclass[a4paper,11pt,onecolumn, accepted=2025-06-03]{quantumarticle}
\pdfoutput=1
\usepackage[utf8]{inputenc}
\usepackage[english]{babel}
\usepackage[T1]{fontenc}
\usepackage{amsmath}
\usepackage{amsthm}
\usepackage{amsfonts}
\usepackage{amssymb}
\usepackage{hyperref}
\usepackage{braket}
\usepackage[numbers,sort&compress]{natbib}

\begin{document}

\title{Low Overhead Qutrit Magic State Distillation}

\author{Shiroman Prakash}
\affiliation{Department of Physics and Computer Science, Dayalbagh Educational Institute, Agra-282005, India}
\email{sprakash@dei.ac.in}

\author{Tanay Saha}
\affiliation{Department of Physics and Computer Science, Dayalbagh Educational Institute, Agra-282005, India}
\affiliation{Present Address: Department of Mathematics, Simon Fraser University, Burnaby, B.C., Canada}
\email{tanays@iitk.ac.in}

\maketitle

\begin{abstract}
We show that using qutrits rather than qubits leads to a substantial reduction in the overhead cost associated with an approach to fault-tolerant quantum computing known as magic state distillation. We construct a family of $[[9m-k, k, 2]]_3$ triorthogonal qutrit error-correcting codes for any positive integers $m$ and $k$ with $k \leq 3m-2$ that are suitable for magic state distillation. In magic state distillation, the number of ancillae required to produce a magic state with target error rate $\epsilon$ is $O(\log^\gamma \epsilon^{-1})$, where the yield parameter $\gamma$ characterizes the overhead cost. 
For $k=3m-2$, our codes have $\gamma = \log_2 (2+\frac{6}{3 m-2})$, which tends to $1$ as $m \to \infty$. 
Moreover, the $[[20,7,2]]_3$ qutrit code that arises from our construction when $m=3$ already has a yield parameter of $1.51$ which outperforms all known qubit triorthogonal codes of size less than a few hundred qubits. 


\end{abstract}

\section{Introduction}

To realize their full potential, quantum computers must achieve large-scale fault tolerance. Magic state distillation \cite{MSD, knill2004faulttolerant} is one of the leading approaches to fault-tolerant quantum computation, in which many noisy magic states are distilled into fewer pure magic states, which are then used to achieve quantum universality via state injection. There has been some encouraging experimental progress in magic state distillation in recent years, e.g., \cite{souza2011experimental, postler2022demonstration, ye2023logical, gupta2024encoding, rodriguez2024experimentaldemonstrationlogicalmagic}. However, the overhead associated with magic state distillation -- i.e., the number of noisy ancillae required to obtain an ancilla in a sufficiently pure magic state -- is high, and represents a significant portion of the cost of quantum computing (as estimated in e.g., \cite{GormanCampbell2017}), though this cost has decreased over the years (e.g., \cite{CampbellHoward2017, litinski2019magic, chamberland2020very, bombin2024fault}). Reducing the overhead of magic state distillation is crucial in order to realize large-scale fault-tolerant quantum computation within the foreseeable future.  At the core of any magic state distillation routine is a quantum error-correcting code with desirable theoretical properties that enable it to distill a few high-fidelity magic states from many low-fidelity magic states -- in this paper, we present a new family of small error-correcting codes based on qutrits that achieve this goal with low overhead.

A magic state distillation routine takes $n$ noisy ancillae as input and, with probability $P$, yields $k$ pure ancillae as output with infidelity suppression $\epsilon_{\text{out}} \sim \epsilon_{\text{in}}^\nu$, where $\nu$ is the noise-suppression exponent. To obtain an ancilla with output error rate $\epsilon_{\rm out}$ less than $\epsilon$, one requires $O\left(\log^\gamma \frac{1}{\epsilon} \right)$ noisy ancillae \cite{MSD, Bravyi_2012}, where 
\begin{equation}
    \gamma = \log_\nu \left( \frac{n}{Pk} \right).
\end{equation}
$\gamma$ is known as the \textit{yield parameter} and characterizes the overhead cost of magic state distillation in the asymptotic limit of many rounds of distillation. 

Of the two magic state distillation routines originally proposed in \cite{MSD}, distillation via the 15-qubit Reed-Muller code  has the lowest overhead, given by  yield parameter of $\gamma= \log_3 15 \approx 2.46$. This code was later shown to be part of a larger family of codes known as \textit{triorthogonal codes} \cite{Bravyi_2012}. For distillation via triorthogonal codes, it can be shown that the probability of successful distillation, $P \to 1$ as $\epsilon_{\rm in} \to 0$, and that $\nu = d$, where $d$ is the distance of the quantum error correcting code. For all other known magic state distillation routines, $P_{\rm success}\leq 1$ and $\nu \leq d$. Therefore, in order to construct distillation routines with the lowest possible yield parameters, it is natural to focus attention on triorthogonal codes. 

Triorthogonal codes have given rise to several reductions to the overhead cost of magic state distillation. In particular, \cite{Bravyi_2012} constructed a family of magic state distillation routines with yield parameter approaching $\gamma= \log_2 3 \approx 1.58$. Triorthogonal codes have been used to construct qubit magic state distillation routines with even lower yield parameters -- but these all seem to require an inordinately large number of qubits. Notably, \cite{Hastings_2018} showed that one can construct qubit triorthogonal codes for which $\gamma<1$, however, the smallest of these codes requires a block size of $2^{58}$ qubits. Other families of qubit distillation routines which achieve a yield parameter that approaches $1$ are known \cite{Jones_2013, Haah_2017, Haah_2018}, however these also seem to require an unwieldy number of qubits.\footnote{After submission of this work for publication, several works\cite{wills2024constantoverheadmagicstatedistillation, golowich2024asymptoticallygoodquantumcodes, nguyen2024goodbinaryquantumcodes} appeared which construct asymptotically good qubit distillation routines with $\gamma \to 0$.} Do there exist low-overhead magic state distillation routines involving a reasonably small number of qubits?  A recent computational search \cite{Nezami_2022} for qubit triorthogonal codes suggest that the answer is no.

An alternative approach to lower the overhead is to construct magic state distillation routines for \textit{qudits} -- quantum systems of prime dimension $p$ -- rather than qubits, which are also of purely theoretical interest due to the direct connection between distillability and contextuality present only for qudits of odd dimension \cite{nature, DPS1, DPS2, Prakash2020contextual}. \cite{ACB} provided the first explicit examples of qudit magic state distillation routines, demonstrating, in particular, that $[[5,1,3]]_3$ code can distill an eigenstate of the qutrit Hadamard gate. However, the distillation routines proposed in \cite{ACB} (which are based on codes that are not triorthogonal) have the disadvantage of having noise suppression exponent $\nu=1$, and therefore infinite yield parameter $\gamma$. Subsequent constructions qudit magic state distillation routines include both triorthogonal codes \cite{CampbellAnwarBrowne,  campbell2014enhanced, Krishna_2019} and non-triorthogonal\footnote{While distillation via non-triorthogonal codes involves higher overheads, such codes do appear to have higher thresholds to noise.} codes \cite{DawkinsHoward,  Howard_2016, 2020golay}. Of these, the lowest known overheads arise from punctured quantum Reed-Solomon codes, which achieve a yield parameter $\gamma = {\log_{\lfloor{(p+1)/3}\rfloor} (p-1)}$ with a single puncture \cite{campbell2014enhanced}, and sublogarithmic yields with multiple punctures \cite{Krishna_2019} for sufficiently large $p$.  While such codes can have modest block sizes $\sim O(p)$, they require qudits of rather unwieldy dimension $p \gtrsim 23$ to obtain $\gamma \sim  1$. 

Given these results, many expect that low-overhead magic distillation, i.e., $\gamma \to 1$, though possible in theory, is not realizable in practice, unless one is willing to work with qudits of large dimension, or codes with large block sizes. Here we show that this expectation is incorrect, by constructing a family of relatively small triorthogonal codes for qutrits that gives rise to a yield parameter approaching $1$.

Our paper is organized as follows. In Section \ref{sec:triorthogonal} we review the construction of triorthogonal codes, which are codes that possess a transversal non-Clifford gate. In Section \ref{sec:our-construction}, we present our new family of qutrit triorthogonal codes. In Section \ref{sec:threshold} (supported by the appendix) we study the noise reduction and compute the threshold of our new family of codes. In Section \ref{sec:overhead} we compute the yield parameter and also estimate the practical overhead cost of our family of codes when only a few rounds of distillation are required. In Section \ref{sec:discussion}, we conclude with discussion and open questions.   

\section{Triorthogonal Codes}
\label{sec:triorthogonal}
We first review the general framework for constructing triorthogonal quantum codes \cite{Bravyi_2012}, adapted to qudits \cite{Krishna_2019} of dimension $p$. First, one must construct a $\kappa$-dimensional self-orthogonal linear subspace $\mathcal T$ of $\mathbb F_p^n$, such that, for any three vectors $h^{(a)},~h^{(b)},~h^{(c)} \in \mathcal T$, the triple product
\begin{equation}
\sum_{i=1}^n h^{(a)}_i h^{(b)}_i h^{(c)}_i =0 \mod p.
\end{equation}
Such a space is known as a ternary \textit{triorthogonal space} of size $n$ and dimension $\kappa$. 

We next puncture\footnote{Recall that puncturing a code at coordinate $j$ means deleting the $j$th column of the generator matrix. Each time a code is punctured, $n$ decreases by $1$.} the triorthogonal space at $k$ different coordinates to form a space generated by a $\kappa\times (n-k)$ dimensional matrix $H$, known as a triorthogonal matrix. We partition $H$ as \begin{equation} H = \begin{pmatrix} H_1 \\ \hline H_0 \end{pmatrix}, \label{triorthogonal-matrix}\end{equation} where $H_0$ consists of those rows of $H$ that are self-orthogonal, and $H_1$ consists of those rows of $H$ that are not. $H$ is called a triorthogonal matrix, and is used to construct a CSS code, known as a triorthogonal code, with stabilizers $\mathcal S_x$ spanned by $H_0$ and $\mathcal S_z$ spanned by  $H^\perp$, and logical operators determined by the rows of $H_1$. Any such code possesses a transversal gate from the third level of the Clifford hierarchy, which, for qutrits, can be taken to be \cite{HowardVala}
\begin{equation}
    T = \sum_k e^{2\pi i k/9}\ket{k}\bra{k}.
\end{equation}
This gate and its generalizations have been studied extensively, see, e.g., \cite{PhysRevA.83.032310, PhysRevA.98.032304, Glaudell_2019, Cui, kalra2024synthesis}.

As a consequence of this transversal non-Clifford gate, a triorthogonal code can be used for magic state distillation of a particular qutrit magic state analogous to the $\ket{H}$ state of \cite{MSD}, (sometimes referred to as an equatorial magic state \cite{jain2020qutrit}), which we denoted as:
\begin{equation}
\ket{M_0} = \sum_j e^{2\pi i j/9}\ket{j}.
\end{equation}
The noise suppression of the resulting routine is, schematically, $\epsilon_{\text{out}} \sim \epsilon_{\text{in}}^d$, where $d$ is the code distance\footnote{Let us point out, when codes that are not triorthogonal are used for magic state distillation, the noise suppression exponent, $\nu$ defined via $\epsilon_{\text{out}} = \epsilon_{\text{in}}^\nu$, is not the same as the code distance $d$.} \cite{MSD, CampbellAnwarBrowne}. Note that, because $H_0$ is self-orthogonal, $\mathcal S_x \subset \mathcal S_z$, and therefore, $\mathcal S_x$ determines the distance of the code. For this reason, it is natural to also demand that the triorthogonal spaces  $\mathcal T$ be \textit{maximal} -- i.e., no additional basis vector can be added to $\mathcal T$ while preserving triorthogonality. The reason for this is as follows. Suppose that there exist two triorthogonal spaces $\mathcal T'$ and $\mathcal T$, such that $\mathcal T' \subset \mathcal T$ -- puncturing $\mathcal T'$ and $\mathcal T$ at the same locations would give rise to stabilizers $\mathcal S_x'$ and $\mathcal S_x$ that satisfy $\mathcal S_x' \subset \mathcal S_x$. Therefore the triorthogonal code obtained from $\mathcal T'$ would have distance $d'\leq d$, the distance of the code obtained from $\mathcal T$.

As an example, we present the well-known family of qubit triorthogonal codes were proposed in \cite{Bravyi_2012}. A representative member of this family correspond to the triorthogonal space $\mathcal T^{(2)}$ generated by the matrix $T^{(2)}$ given by
\begin{equation}
    T = \begin{pmatrix}
        1 & 1 & 1 & 1 & 0 & 0 & 0 & 0 & 1 & 1 & 1 & 1 & 0 & 0 & 0 & 0 \\
        0 & 0 & 0 & 0 & 1 & 1 & 1 & 1 & 1 & 1 & 1 & 1 & 0 & 0 & 0 & 0 \\
        0 & 1 & 0 & 1 & 0 & 1 & 0 & 1 & 0 & 1 & 0 & 1 & 0 & 1 & 0 & 1 \\
        0 & 0 & 1 & 1 & 0 & 0 & 1 & 1 & 0 & 0 & 1 & 1 & 0 & 0 & 1 & 1 \\
        0 & 0 & 0 & 0 & 0 & 0 & 0 & 0 & 1 & 1 & 1 & 1 & 1 & 1 & 1 & 1 
    \end{pmatrix}.
\end{equation}
This can be punctured at coordinates $1$ and $5$ to give rise to a triorthogonal matrix of the form:
\begin{equation}
    H = \begin{pmatrix} H_1 \\ \hline H_0 \end{pmatrix} = \begin{pmatrix}
        & 1 & 1 & 1 & 0 & 0 & 0 & 1 & 1 & 1 & 1 & 0 & 0 & 0 & 0 \\
        & 0 & 0 & 0 & 1 & 1 & 1 & 1 & 1 & 1 & 1 & 0 & 0 & 0 & 0 \\
        \hline 
        & 1 & 0 & 1 & 1 & 0 & 1 & 0 & 1 & 0 & 1 & 0 & 1 & 0 & 1 \\
        & 0 & 1 & 1 & 0 & 1 & 1 & 0 & 0 & 1 & 1 & 0 & 0 & 1 & 1 \\
        & 0 & 0 & 0 & 0 & 0 & 0 & 1 & 1 & 1 & 1 & 1 & 1 & 1 & 1 
    \end{pmatrix},
\end{equation}
that defines a $[[14,2,2]]_2$ triorthogonal code with yield parameter $\gamma = \log_2 7 \approx 2.81$. 

\section{Our Construction}
\label{sec:our-construction}
We define a triorthogonal space denoted as $\mathcal T_m$ of size $n=9m$ and dimension $3m$, spanned by basis vectors $w$ and $\{v^{(1)},~v^{(2)},~\ldots v^{(3m-1)} \}$ given by 
\begin{equation}
    w=(0,~1,~2, 0,~1,~2, \ldots)
\end{equation} 
and
\begin{equation}
    v^{(a)}_i = \begin{cases}
                0 & 0 < i \leq 3(a-1), \\
                1 & 3(a-1) < i \leq 3a, \\
                0 & 3a < i \leq  n-3, \\
                2 & n-3 < i \leq n.
                \end{cases}
\end{equation}
By direct computation, it is easy to check that $\mathcal T_m$ is triorthogonal and maximal.

Explicitly, for $m=2$, the generator matrix for $\mathcal T_m$ is given by
\begin{equation}
\begin{pmatrix}
    0 & 1 & 2 & 0 & 1 & 2 & 0 & 1 & 2 & 0 & 1 & 2 & 0 & 1 & 2 & 0 & 1 & 2 \\
    1 & 1 & 1 & 0 & 0 & 0 & 0 & 0 & 0 & 0 & 0 & 0 & 0 & 0 & 0 & 2 & 2 & 2 \\
    0 & 0 & 0 & 1 & 1 & 1 & 0 & 0 & 0 & 0 & 0 & 0 & 0 & 0 & 0 & 2 & 2 & 2 \\
    0 & 0 & 0 & 0 & 0 & 0 & 1 & 1 & 1 & 0 & 0 & 0 & 0 & 0 & 0 & 2 & 2 & 2 \\
    0 & 0 & 0 & 0 & 0 & 0 & 0 & 0 & 0 & 1 & 1 & 1 & 0 & 0 & 0 & 2 & 2 & 2 \\
    0 & 0 & 0 & 0 & 0 & 0 & 0 & 0 & 0 & 0 & 0 & 0 & 1 & 1 & 1 & 2 & 2 & 2
\end{pmatrix}.
\end{equation}

The 9-qutrit triorthogonal space that arises from this construction when $m=1$ coincides with a linear Reed-Muller code in two variables, which can be punctured to give rise to the $[[8,1,2]]_3$ code used for distillation in \cite{CampbellAnwarBrowne}. The 27-qutrit triorthogonal space that arises from this construction when $m=3$ contains the linear Reed-Muller code in three variables as a proper subspace.

One can puncture $\mathcal T_m$ once per block in any $3m-2$ blocks of three trits to obtain a $[[9m-k,k,2]]_3$ CSS code for $k \leq 3m-2$. Explicitly, let us choose to puncture at coordinates $3j+1$, for all $0 \leq j < 3m-2$. For $m=2$, this leads to a triorthogonal matrix $H$ given by
\begin{equation}
H= \begin{pmatrix} H_1 \\ \hline H_0 \end{pmatrix} = \begin{pmatrix}
    1 & 1 & 0 & 0 & 0 & 0 & 0 & 0 & 0 & 0 & 0 & 2 & 2 & 2 \\
    0 & 0 & 1 & 1 & 0 & 0 & 0 & 0 & 0 & 0 & 0 & 2 & 2 & 2 \\
    0 & 0 & 0 & 0 & 1 & 1 & 0 & 0 & 0 & 0 & 0 & 2 & 2 & 2 \\
    0 & 0 & 0 & 0 & 0 & 0 & 1 & 1 & 0 & 0 & 0 & 2 & 2 & 2 \\
    \hline 
    1 & 2 & 1 & 2 & 1 & 2 & 1 & 2 & 0 & 1 & 2 & 0 & 1 & 2 \\
    0 & 0 & 0 & 0 & 0 & 0 & 0 & 0 & 1 & 1 & 1 & 2 & 2 & 2
\end{pmatrix}.
\end{equation}
For any $m$, the two stabilizers in $\mathcal S_x$ will be of the form $(1,~2,~1,~2,~\ldots,~0,~1,~2,~0,~1,~2)$ and $(0, \ldots,~0, ~1,~1,~1,~2,~2,~2)$, which allow us to detect, but not correct, a single $Z$-error, leading us to conclude that the code has distance $2$.

\section{Threshold}
\label{sec:threshold}
Let us first compute the threshold to noise for these distillation routines. Note that, although the threshold does not affect the computation of the yield parameter $\gamma$, it does have a direct impact on the actual overhead cost of distillation in the limit when only a few rounds of distillation are used, as is discussed in the next section.

To describe a general mixed qutrit state, one requires 8 independent noise parameters. However, after twirling, as described in \cite{CampbellAnwarBrowne} and \cite{jain2020qutrit}, noisy qutrit $\ket{M_0}$ magic states can be expressed in terms of only two noise parameters, $\epsilon_1$ and $\epsilon_2$, via 
\begin{equation}
    \rho(\epsilon_1,\epsilon_2) = (1-\epsilon_1 -\epsilon_2) \ket{M_0}\bra{M_0} + \epsilon_1 \ket{M_1}\bra{M_1} + \epsilon_2 \ket{M_2}\bra{M_2},
\end{equation}
where $\ket{M_j}=Z^j\ket{M_0}$. The performance of a qutrit magic state distillation routine is thus characterized by two functions $\epsilon_1'(\epsilon_1, \epsilon_2)$ and $\epsilon_2'(\epsilon_1, \epsilon_2)$, which determine a basin of attraction in the space of all convex combinations of the $\ket{M_i}\bra{M_i}$ that distill to the desired magic state.  

\begin{figure}[h!]
    \centering
    \includegraphics[width=0.6\textwidth]{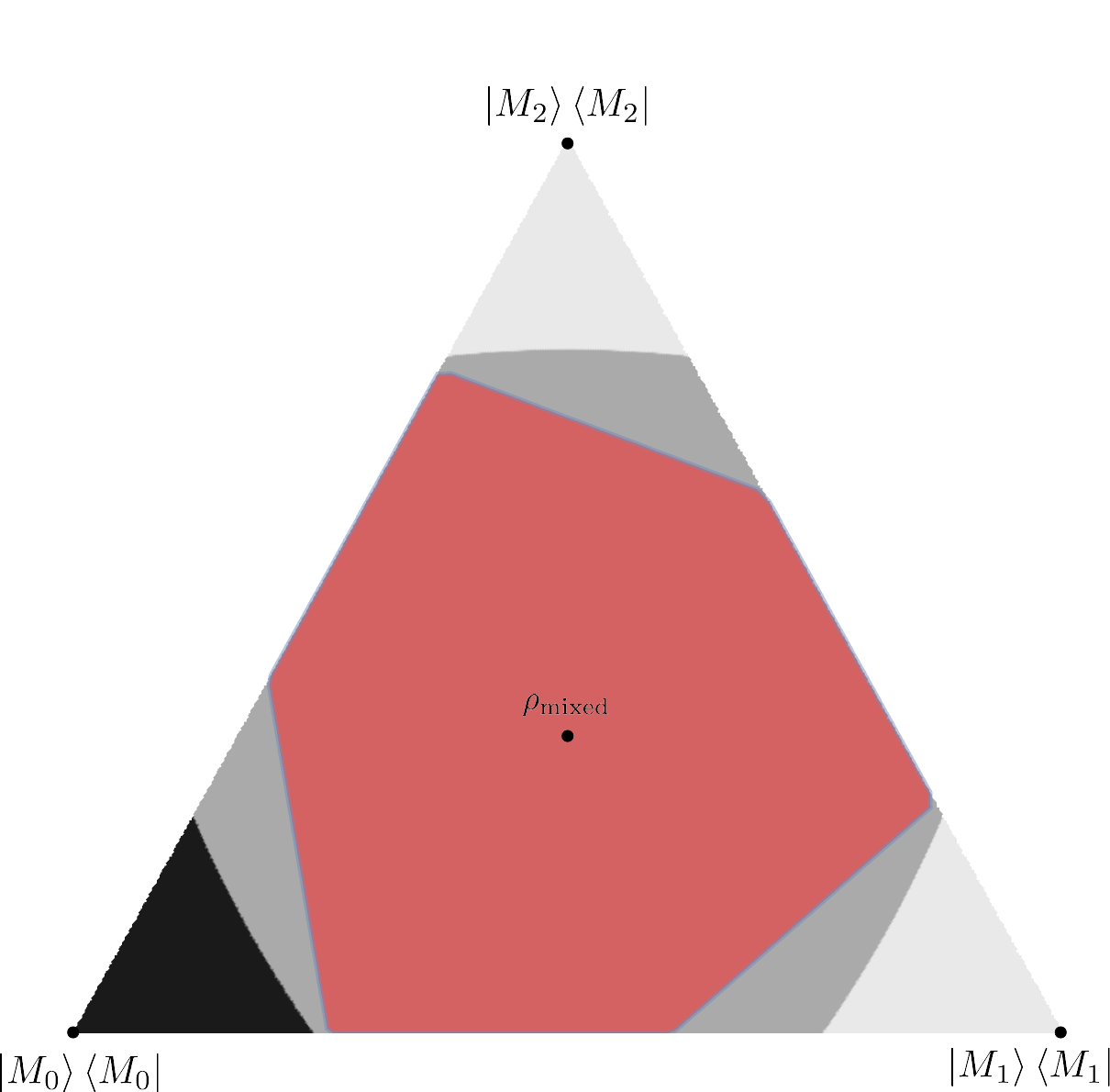}
    \caption{Any noisy magic state can be twirled to lie in the triangle spanned by convex combinations of $\ket{M_i}$, pictured above. The region of state space that distills to the magic state $\ket{M_0}$ via the $[[17,1,2]]_3$ code is shown in black. The light gray regions distill to magic states $\ket{M_1}$ and $\ket{M_2}$. The red and dark gray regions distill to the maximally mixed state, at the centre of the triangle. (The red region is inside the Wigner polytope, and cannot be distilled to a useful magic state by any code.)}
    \label{fig:distillable-region}
\end{figure}

To obtain the highest thresholds (i.e., largest basins of attraction) possible using our construction, we set $k=1$. We computed both $\epsilon_1'(\epsilon_1, \epsilon_2)$ and $\epsilon_2'(\epsilon_1, \epsilon_2)$ for small values of $m$ following \cite{CampbellAnwarBrowne}, and found that $m=2$ gives the largest basin of attraction. Details of this computation are provided in Appendix \ref{app:we}. Explicitly, up to order, $\epsilon_i^3$, we found, for $m=2$,
\begin{eqnarray}
    \epsilon_1'(\epsilon_1, \epsilon_2) = \epsilon _1^2+ 2 \left( \epsilon _1^3+ \epsilon _2 \epsilon _1^2+15 \epsilon _2^2 \epsilon _1 \right) + O(\epsilon_i^4)\\
    \epsilon_2'(\epsilon_1, \epsilon_2) = \epsilon _2^2+2 \left(\epsilon _2^3+\epsilon _1 \epsilon _2^2+15 \epsilon _1^2 \epsilon _2\right)+O(\epsilon_i^4).
\end{eqnarray}
We plot the basins of attraction of the $[[17,1,2]]_3$ code in the two-dimensional space of qutrit state space in Figure \ref{fig:distillable-region}. As can be seen from this figure, the  region that distills to $\ket{M_0}$ extends fairly close to the Wigner polytope \cite{Veitch_2012, Veitch_2014} which sets the theoretical bound for the best possible magic state distillation routine, and is tied to the onset of (state-dependent) contextuality \cite{nature, Prakash2020contextual}.

To condense the information contained in the basin of attraction plotted in Figure \ref{fig:distillable-region}, it is often customary to assume depolarizing noise, by demanding that $\epsilon_1=\epsilon_2=\delta/3$. This allows us to  characterize  the  distillation performance by a single function $\delta'(\delta)$, and the basin of attraction by a single inequality, $\delta<\delta_*$, where $\delta_*$ is the threshold to depolarizing noise. We found that, for $m\geq 2$ and $k=1$,
\begin{equation}
\delta' = \frac{1}{3}\delta^2+\frac{8 + 9 (m-1)}{9} 2\delta^3+ \frac{41 + 54 (m-1)}{27} 2\delta^4 +O(\delta^5).
\end{equation}

\begin{figure}
    \centering
    \includegraphics[width=0.8\textwidth]{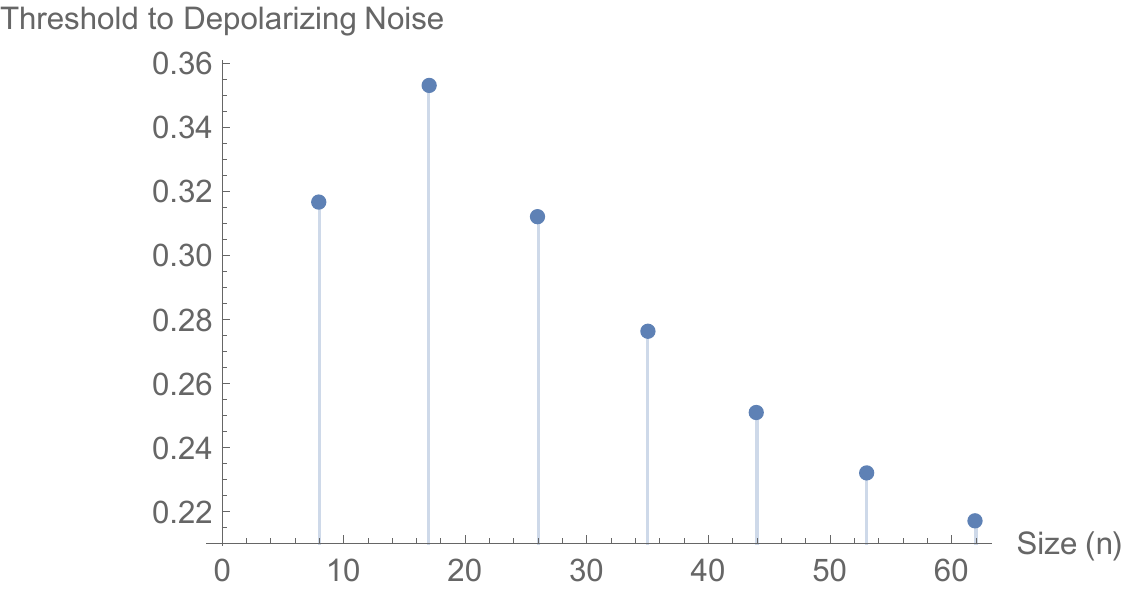}
    \caption{The threshold to depolarizing noise for distillation via the $[[9m-1,1,2]]_3$ codes, as a function of code size. The $[[17,1,2]]_3$ code has the highest threshold, $\delta_*=0.353$.}
    \label{fig:depolarizing-noise-thresholds}
\end{figure}

The thresholds to depolarizing noise for codes with $k=1$ and $m<8$ are plotted in Figure \ref{fig:depolarizing-noise-thresholds}. We find that the $[[17,1,2]]_3$ code obtained from our construction when $m=2$ and $k=1$ has the highest depolarizing noise threshold, $\delta_*=0.353$, exceeding that of the $[[8,1,2]]_3$ punctured Reed-Muller code of \cite{CampbellAnwarBrowne}, whose noise threshold is $\delta_*=0.317$.\footnote{For distillation of certain other qutrit magic states, slightly higher thresholds are possible using the 11-qutrit Golay code \cite{2020golay} or \cite{DawkinsHoward}, but at a cost of a much lower yield.} The bound for the best possible threshold to depolarizing noise determined by the Wigner polytope is $\delta_*^{\text{max}}=0.467$.

Let us also discuss the case of codes with $k>1$, restricting to the case of depolarizing noise for simplicity. We find that, for the $[[9m-k, k,2]]_3$ code, each individual output qutrit has depolarizing noise $\delta'$ given by 
\begin{equation}
    \delta' = \frac{A_2(k,m)}{6}\delta^2+O(\delta^3) \label{noise-reduction-eq}
\end{equation}
where
\begin{equation}
    A_2 = \begin{cases} 2k^2 & k< 3m-2 \\ 2k^2+6 & k= 3m-2 \end{cases}
\end{equation}
is the number of logical $Z$ or $Z^2$ operators of weight $2$ that act on any one individual output qutrit. For the case of $k=3m-2$, $A_2$ is computed in the Appendix; the case $k<3m-2$ can be determined in a similar manner. We see that the threshold for large $k$ is $\delta_* \approx \frac{6}{A_2} \sim \frac{3}{k^2}$. The threshold decreases quadratically with $1/k$ rather than linearly, as it does for the case of qubits \cite{Bravyi_2012}. 

The success probability is given by \begin{equation}P_s(k,m)=1 - \frac{18m-2k}{3}\delta+O(\delta^2). \label{success-probability}\end{equation}

\section{Overhead}
\label{sec:overhead}
How should one measure the overhead of a magic state distillation protocol? For both qubits and qudits, Magic state distillation, as proposed in \cite{MSD, ACB, CampbellAnwarBrowne}, starts from the assumption that one already has a fault-tolerant quantum computer able to implement Clifford unitaries and prepare/measure stabilizer states without any error\footnote{Qudit generalizations of the color code and surface code that allow for fault-tolerant implementations of the qudit Clifford group are discussed in \cite{Anwar_2014, Watson_2015, Watson_2015-2}.}. Following \cite{MSD, Bravyi_2012}, we define the (theoretical) overhead cost of magic state distillation simply as the number of low-fidelity magic states needed to produce a high-fidelity magic state, assuming free and perfect Clifford operations. Explicitly, if $N_{\rm in}$ with noise rate $\delta_{\rm in}$ are needed to produce $N_{\rm out}$ magic states with noise rate $\delta_{\rm out}$, following \cite{meier2012magicstatedistillationfourqubitcode, Bravyi_2012, Jones_2013}, we define the cost of distillation to be 
\begin{equation}
    C(\delta_{\rm in}, \delta_{\rm out})=\frac{N_{\rm in}}{N_{\rm out}}. \label{distillation-cost}
\end{equation}  

Is it possible to compare fault-tolerant schemes for qutrits to qubits? In general there are more noise channels for qutrits, and a noisy qutrit may be harder to produce than a noisy qubit. One might model these differences by introducing ``conversion constants'' $K>1$ and $K'>1$, such that $n_{\rm qutrit} = K n_{\rm qubit}$ and $\epsilon_{\rm qutrit} = K' \epsilon_{\rm qubit}$. However, the yield parameter $\gamma$ measures how the number of noisy magic states scales with the logarithm of the target error rate in the asymptotic limit of many rounds of distillation, and is clearly independent of $K$ and $K'$ -- therefore it appears meaningful to compare yield parameters across qudits of different dimensionalities. In subsection \ref{sec:yield-1} we compute the yield parameter for our codes and compare to yield parameters from similar qubit codes.

In practice, however, only a few rounds of distillation may be needed, so the yield parameter is not necessarily meaningful. We study the distillation cost as a function of target noise rate in section \ref{sec:practical}. 

The distillation cost is a very natural measure of overhead cost associated with magic state distillation -- indeed, the entire resource theory of magic (e.g., \cite{Veitch_2012, Mari_2012, Veitch_2014, nature, Howard_2017, Wang_2020, Koukoulekidis_2022}) seeks to place theoretical bounds on this measure. However, in practice, the cost of fault-tolerant implementation of the Clifford group must also be taken into consideration when estimating the total overhead cost of quantum computing via magic states, and this requires a more careful study of the circuits used for implementing the distillation protocol. While we expect that distillation with codes with low theoretical overhead cost may give rise to lower overheads in practice, a detailed estimate of the total overhead cost of fault-tolerant quantum computation via magic state distillation using the new family of qutrit error-correcting codes proposed here is beyond the scope of the present work.\footnote{Indeed, a variety of creative techniques and optimizations have been proposed in the literature for magic state preparation and distillation that can greatly reduce the total overhead cost in practice. Other techniques, such as combining magic state distillation with gate synthesis \cite{CampbellHoward2017} can further reduce the overhead cost for fault-tolerant implementation of a given algorithm. All these ideas can, presumably, also be applied to qutrits, but the details have not been worked out in the literature.}

\subsection{Yield Parameter}
\label{sec:yield-1}
Let us first compute the yield parameter, which, captures the overhead cost in the limit of many rounds of distillation.

\begin{figure}
    \centering
    \includegraphics[width=0.8\textwidth]{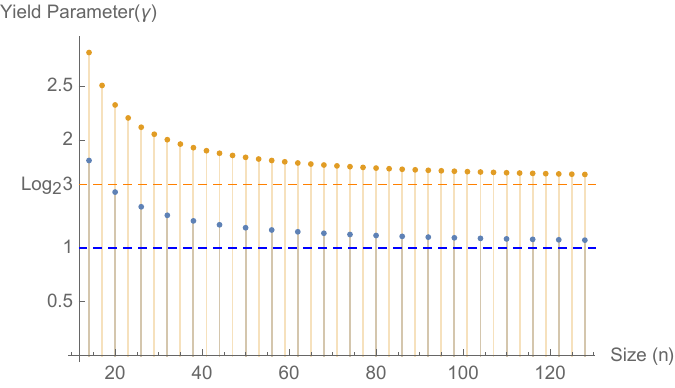}
    \caption{Yield parameters $\gamma$ of the lowest overhead $[[n,k,2]]_3$ triorthogonal code produced by our construction are shown in blue, as a function of code size $n$. As $n\to \infty$, $\gamma \to 1$, shown by the blue line. For comparison, yield parameters for binary $[[n,k,2]]_2$ triorthogonal codes in \cite{Bravyi_2012} with the same $n$ are plotted in orange on the same graph -- for these codes $\gamma \to \log_2 3$ as $n \to \infty$, shown by the orange dashed line.}
    \label{fig:yield}
\end{figure}

To obtain triorthogonal codes with the maximal yield parameter, we set $k=3m-2$, which is the maximum value for which the code has distance $2$. The resulting $[[6m+2,3m-2,2]]_3$ triorthogonal codes have yield parameter given by,
\begin{equation}
    \gamma = \log_2 (2+\frac{6}{3 m-2}),
\end{equation}
which approaches $1$ as $m \to \infty$. The yield parameter as a function of code size $n$ is plotted in Figure \ref{fig:yield}. The advantages over qubit codes of comparable size is striking. Let us explicitly list the yields for the smallest codes in our construction, obtained by setting $k=3m-2$.
\begin{itemize}
    \item For $m=2$, we obtain a $[[14,4,2]]_3$ code with a yield parameter of $1.81$, which is better than the best known qubit triorthogonal codes of size less than 50 \cite{Bravyi_2012, Nezami_2022}. 
    \item For $m = 3$, we obtain a $[[20,7,2]]_3$ code, whose yield parameter is $1.51 < \log_2 3$, outperforming all the codes constructed in \cite{Bravyi_2012}, which, to our knowledge, are the best known qubit triorthogonal codes with block size less than a few hundred qubits. 
    \item For $m = 4$, we obtain a $[[26,10,2]]_3$ code, whose yield parameter is $1.38$.
    \item For $m = 5$, we obtain a $[[32,13,2]]_3$ code, whose yield parameter is $1.30$.
    \item For $m = 6$, we obtain a $[[38,16,2]]_3$ code, whose yield parameter is $1.25$.
    \item For $m = 7$, we obtain a $[[44,19,2]]_3$ code, whose yield parameter is $1.21$.
    \item For $m=8$, we obtain a $[[50, 22,2]]_3$ code with yield parameter $1.18$.  To the best of our knowledge, to get a comparable yield parameter, one requires a code with $\sim 1000$ qubits, using  triorthogonal codes constructed from punctured binary Reed-Muller codes in \cite{Hastings_2018, Haah_2018}.
\end{itemize} 
Note also that previous constructions of qudit-based triorthogonal codes in \cite{campbell2014enhanced, Krishna_2019} require qudits of dimension at least $17$ to obtain comparable improvements in yields.

\subsection{Practical Estimates}
\label{sec:practical}
In this section, we study the cost of magic state distillation when only a few rounds of distillation are needed, which is, perhaps, more relevant in practice than the yield parameter $\gamma$.

To see that our codes may still offer an advantage, let us compare the distillation cost using our family of codes to distillation using the $[[15,1,3]]_2$ code of \cite{MSD}. Of course, it is not clear how to fairly compare qubit noise to qudit noise -- following \cite{campbell2014enhanced} for simplicity, we work with the depolarizing noise $\delta$. Expressed in terms of depolarizing noise, the noise reduction of the $[[15,1,3]]_2$ code for distillation is 
\begin{equation}
 \delta' = \frac{35}{4} \delta^3 + O(\delta^4).
\end{equation}
The probability of successful projection is $P_s=1-\frac{15}{2}\delta + O(\delta^2)$.

Using the $[[15,1,3]]_2$ code, starting with a $\delta_{\rm in}=.001$, a single round of distillation would give $\delta_{\rm out} \approx 10^{-8}$, with a distillation cost of $15.2$. Now consider distillation with the $[[32,13,2]]_3$ code from our family of codes. Starting with $\delta_{\rm in} =.001$, after three rounds of distillation with the $[[32,13,2]]_3$ code,  using Eq. \eqref{noise-reduction-eq} one finds, $\delta_{\rm  out} \approx \left(\frac{{A}_2(5,13)}{6}\right)^7 \delta_{\rm in}^8 \approx 2\times 10^{-12}$,  with a distillation cost of $C \approx \left( \frac{32}{13}\right)^3/P_s(.001)\approx 15.2$. Although depolarizing noise for qubits may not be directly comparable to depolarizing noise for qutrits, for the same distillation cost, qutrits with $\delta \sim 10^{-12}$ are presumably preferable to qubits with $\delta \sim 10^{-8}$.   

One also has the freedom to choose a different code (i.e., a code with a different value of $m$) during each round, as described in \cite{meier2012magicstatedistillationfourqubitcode, Bravyi_2012}. Suppose we distill with an $[[n_1,k_1,2]]_3$ code on round $1$, an $[[n_2,k_2,2]]_3$ on round $2$ and an  $[[n_3,k_3,2]]_3$ on round $3$. Assuming $\delta_{\rm in} \ll A_2/6$, we have,
\begin{equation}
    \delta_{\rm out} \approx {A}_2(k_3) {A}_2(k_2)^2 {A}_2(k_1)^4 \frac{\delta_{\rm in}^8}{6^7}, 
\end{equation}
with cost of distillation
\begin{equation}
    C \approx \left(\frac{n_3}{k_3}\right) \left(\frac{n_2}{k_2}\right) \left(\frac{n_1}{k_1}\right).
\end{equation}
Because triorthogonal codes with larger $k$ typically have better rates, but smaller thresholds, there is some optimization required in the choice of distillation sequence. Heuristically, it appears advantageous to use distillation protocols with smaller $k$ for early rounds of distillation, and protocols with larger $k$ for later rounds.    

\begin{figure}
    \centering
    \includegraphics[width=0.8\linewidth]{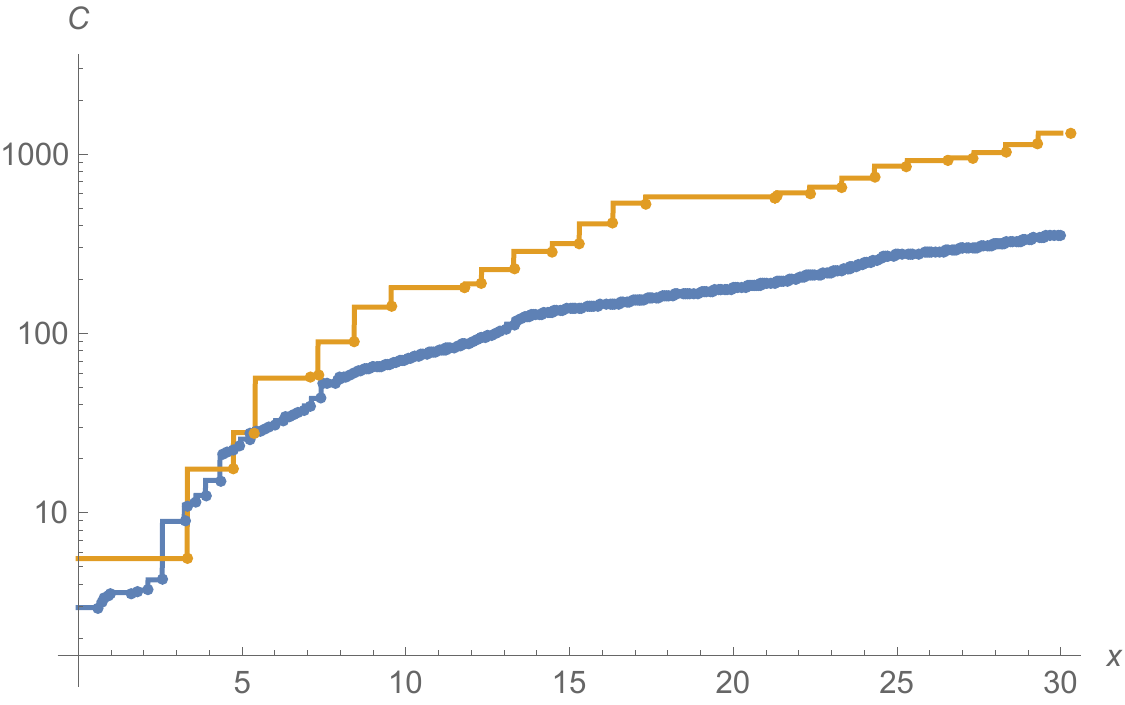}
    \caption{We plot the distillation cost $C=\frac{N_{\rm in}}{N_{\rm out}}$ needed to produce magic states a target depolaring noise rate of $\delta_{\rm out}=10^{-x}$, starting from $\delta_{\rm in} =0.02$ for optimal distillation sequences of from our family of qutrit triorthogonal codes in blue. Though a comparison may not be fair, we also plot, in orange, similar results for optimal sequences of small qubit triorthogonal codes ($n\leq 128$), taken from \cite{Bravyi_2012}. (These optimal sequences typically involve one round of distillation with the $[[15,1,3]]_2$ code, followed by additional rounds with other triorthogonal codes with higher rates.)}
\label{fig:distillation-cost}
\end{figure}

Using a randomized search we determined optimal sequences for 
qutrit distillation routines using our family of codes. (The optimal sequences were computed subject to the following simplifying assumptions: for any $m$, we always choose $k=k_{\rm max}$; we restricted to codes with $m\leq 12$; we restricted the total rounds of distillation to be 5 or less.) The results are
shown in Figure \ref{fig:distillation-cost}, which presents the optimal cost $C$ as a function of target depolarizing noise rate $\delta_{\rm target}=10^{-x}$.  For comparison, we plot distillation costs arising from optimal sequences of distillation using small qubit triorthogonal codes (including $[[15,1,3]]_2$ code, the $[[10,2,2]]_2$ code of \cite{meier2012magicstatedistillationfourqubitcode} and the block $[[3k+8,k,2]]_2$ codes in \cite{Bravyi_2012}), which were computed in \cite{Bravyi_2012} for $\delta_{\rm in}=0.02$.  The plot suggests an advantage in overhead costs persists for qutrits for target noise rates better than $10^{-10}$. 

Do these advantages hold up in practice, when one takes into account the cost and imperfections of Clifford operations? While we leave this for future work, let us make a few comments about it below. The $[[9m-k,k,2]]_3$ qutrit triorthogonal codes constructed in this paper offer similar noise reduction to $[[3k+8,k,2]]_2$ block triorthogonal codes of \cite{Bravyi_2012} but have better rates. Overhead costs of distillation with the $[[3k+8,k,2]]_2$ block codes have been compared to the $[[15,1,3]]_2$ code, with mixed results \cite{fowler2013surface, GormanCampbell2017}. At least in noise regimes where distillation via $[[3k+8,k,2]]_2$ block codes is favored, our qutrit codes have a chance to give rise to lower overhead costs in practice, as evidenced by the computations in this section. However, for distillation with qutrits, no small triorthogonal code with distance $3$, analogous to the $[[15,1,3]]_2$ code, is known at present. Therefore, unless such a code is discovered, it appears that, in order to distill qutrit magic states with low overhead, one must make do with the family of codes presented in this paper.

\section{Discussion}
\label{sec:discussion}

Several authors \cite{CampbellAnwarBrowne, campbell2014enhanced, Krishna_2019} have noticed theoretical advantages in the overhead cost of magic state distillation with qudits rather than qubits. However, none of these constructions provided any overhead reductions for qutrits. Indeed, prior to this work, the only practically useful code for qutrit distillation was the $[[8,1,2]]_3$ code of \cite{CampbellAnwarBrowne} -- which does not appear to offer any substantial advantage over the $[[15,1,3]]_2$ qubit code of \cite{MSD}.\footnote{Other qutrit distillation protocols, such as those of \cite{ACB, DawkinsHoward, 2020golay}, while theoretically important, give rise to distillation routines with impractically high overhead costs.} We provided a simple construction of a new family of qutrit triorthogonal codes that have yield parameter $\gamma \to 1$ in the limit $m \to \infty$. The yield parameters of even the smallest codes in our family, involving as few as 20 qutrits, are substantially better than all known constructions involving hundreds of qubits or fewer. Even when only a few rounds of distillation are needed, these codes can outperform the $[[15,1,3]]_2$ code, or codes of \cite{Bravyi_2012}, in terms of overhead cost, assuming perfect and free Clifford unitaries and stabilizer measurements.

This work indicates that the space of qutrit error-correcting codes remains relatively unexplored. Another natural question is whether there exist other qutrit triorthogonal codes with better yields, or higher distances? In particular, what is the smallest qutrit triorthogonal code with distance $3$? We carried out an preliminary computer search over triorthogonal spaces of small dimension. The only non-trivial triorthogonal spaces of size $n\leq 21$ and dimension $\kappa \geq \lfloor n/3\rfloor$ correspond to the spaces produced by our construction with $m=1$ and $m=2$. We therefore believe that the codes presented in this paper are somewhat distinguished, at least amongst very small qutrit triorthogonal codes.

The existence of the family of codes presented here demonstrates a concrete theoretical advantage to pursuing fault-tolerant quantum computation with qutrits rather than qubits. A number of promising experimental realizations of qutrit-based quantum computers exist (e.g., \cite{Lanyon2007ManipulatingBQ, Bianchetti_2010, yurtalan2020characterization, kononenko2020characterization, Ringbauer2021AUQ, Goss_2022, Schutz:2022ydy, Lindon2023, goss2023extending, subramanian2023efficient}), and if one has already made the decision to work with qutrits, the magic state distillation routines presented in this paper appear to be the best currently available. However, if one wishes to assess whether or not the possible advantages for large-scale fault-tolerance in the long term outweigh the increased complexities associated with control of qutrits in the short-term, more work needs to be done. In particular, it is important to know how well do these theoretical advantages hold up in practice, when the cost of (imperfect) Clifford unitaries is taken into account. 

\section*{Acknowledgements}
SP thanks Prof. P.S. Satsangi for inspiration, guidance, and emphasizing to him the importance of qudits of odd prime dimension. The authors also thank Noah Goss, Prahladh Harsha, Rishabh Kothary, Jeongwan Haah and Luca Giuzzi for discussions on related topics. Some of the computations in this paper were performed with assistance of the software package MAGMA \cite{magma}.
The authors acknowledge the support of MeitY QCAL, Amazon Braket and DST-SERB grant (CRG/2021/009137).

\appendix
\section{Threshold Computations}
\label{app:we}
\subsection{General Analysis}
In this Appendix, we directly extend the analysis of section 4 of \cite{Bravyi_2012} to qutrits. A similar analysis, restricted to $k=1$, appears in \cite{CampbellAnwarBrowne}.

The magic state we wish to distill is $\ket{M_0}=T \ket{+}$, where $\ket{+}=\frac{1}{\sqrt{3}}(\ket{0}+\ket{1}+\ket{2})$ and $T$ is the qudit version of the $\pi/8$ gate defined in \cite{HowardVala, CampbellAnwarBrowne}. A triorthogonal code is a CSS-code with a transversal $T$ gate, constructed from a triorthogonal matrix, of the form given in Eq. \eqref{triorthogonal-matrix}. Let $\mathcal H_0 = {\rm span}(H_0)$ and $\mathcal H= {\rm span}(H)$. The $X$-stabilizers are given by $\mathcal S_x=\mathcal H_0$ and the $Z$-stabilizers are given by $\mathcal H^\perp$. The rows of $H_1$, which we denote by $f^{(a)}$ for $a \in \{ 1, \ldots, k\}$, define the logical $Z$ operators of the code as in \cite{Bravyi_2012} via
\begin{equation}
\bar{Z}^{(a)} = Z^{f^{(a)}_1}\otimes Z^{f_2^{(a)}} \otimes \ldots \otimes Z^{f^{(a)}_n}.
\end{equation}

 A twirled noisy magic state can be written in the form,
\begin{equation}
    \rho(\epsilon_1,\epsilon_2) = (1-\epsilon_1 -\epsilon_2) \ket{M_0}\bra{M_0} + \epsilon_1 \ket{M_1}\bra{M_1} + \epsilon_2 \ket{M_2}\bra{M_2},
\end{equation}
where $\ket{M_j}=Z^j\ket{M_0}$. As in the case of qubits, all noise is in the form of $Z$ errors on input qutrits, which translate into $Z$ errors on the output magic state. However, unlike the case of qubits, both $Z$ and $Z^2$ errors may occur with \textit{a priori} independent probabilities -- $\epsilon_1$ is the probability of a $Z$ error and $\epsilon_2$ is the probability of a $Z^2$ error. For simplicity, it is often convenient to assume depolarizing noise, in which case we can set $\epsilon_1=\epsilon_2=\frac{\delta}{3}$.

The performance of the magic state distillation routine via a triorthogonal code can be expressed in terms of weight enumerators $\mathcal S_x$ and $\mathcal S_z$ and their duals. However, unlike the case of qubits, one must work with complete weight enumerators (e.g., \cite{macwilliams77}), rather than simple weight enumerators. To this end, given a ternary vector $u \in F_3^n$ we define $|u|_1$ to be the number of entries of $u$ equal to $1$, and $|u|_2$ to be the number of entries of $u$ equal to $2$. The Hamming weight of $u$ is $|u|=|u|_1+|u|_2$. For example, if $u=(1,2,2,0,0,0,2,2)$, $|u|_1=1$, $|u|_2=4$ and $|u|=5$. 

Projecting $\rho(\epsilon_1,\epsilon_2)^{\otimes n}$ onto the codespace of a triorthogonal code is successful with probability,
\begin{equation}
    P_s = \sum_{u \in \mathcal H_0^\perp} (1-\epsilon_1-\epsilon_2)^{n-|u|_1-|u|_2}\epsilon_1^{|u|_1}\epsilon_2^{|u|_2}. 
\end{equation}
If we restrict to depolarizing noise, we have
\begin{equation}
    P_s \sum_{u \in \mathcal H_0^\perp}  (1-\frac{2\delta}{3})^{n-|u|}\left(\frac{\delta}{3} \right)^{|u|} = \frac{1}{|\mathcal H_0|}\sum_{f \in \mathcal H_0}  \left(1-\delta \right)^{|u|},
\end{equation}
where we used the MacWilliams identity for the simple weight enumerator of a classical ternary code \cite{macwilliams77} in the last equality. 

Our code encodes $k$ logical qutrits. After projection and decoding, the output state is 
\begin{equation}
    \rho_{\rm out} = \sum_{x \in F_3^k} P_{\rm out}(x)\ket{M_{x}}\bra{M_{x}}
\end{equation}
where $\ket{M_x}=\ket{M_{x_1}} \otimes \ket{M_{x_2}} \otimes \ldots \otimes \ket{M_{x_k}},$ and
\begin{equation}
    P_{\rm out}(x) = \sum_{f \in \mathcal S_Z + x_1 f^{(1)} + \ldots x_k f^{(k)}} (1-\epsilon_1-\epsilon_2)^{1-|f|}\epsilon_1^{|f|_1} \epsilon_2^{|f|_2}.
\end{equation}
The reduced density matrix describing the $a$th output qutrit can be written as,
\begin{equation}
    \rho = (1-\bar{\epsilon}_1^{(a)}-\bar{\epsilon}_2^{(a)}) \ket{M_0}\bra{M_0}+\bar{\epsilon}_1^{(a)} \ket{M_1}\bra{M_1}+\bar{\epsilon}_2^{(a)} \ket{M_2}\bra{M_2}
\end{equation}
where
\begin{equation}
    \bar{\epsilon}_j^{(a)} = \frac{1}{P_s} \sum_{x: ~x_a = j} P_{\rm out}(x).
\end{equation}

If we restrict to depolarizing noise, demanding that $\bar{\epsilon}_1^{(a)}=\bar{\epsilon}_2^{(a)} = \frac{\bar{\delta}^{(a)}}{3}$, then one can write,
\begin{equation}\begin{split}
    \frac{2}{3}\bar{\delta}^{(a)} & = 1 - \frac{1}{P_s} \sum_{x: ~x_a = 0} P_{\rm out}(x). \\
    & = 1 - \frac{\sum_{u \in \mathcal (H_0 \oplus f^{(a)})^\perp} (1-\frac{2\delta}{3})^{n-|u|}\left(\frac{\delta}{3} \right)^{|u|} }{\sum_{u \in H_0^\perp}  (1-\frac{2\delta}{3})^{n-|u|}\left(\frac{\delta}{3} \right)^{|u|} }
    \end{split}. \label{error-output-qutrit}
\end{equation}
Note that, again using the ternary MacWilliams identity, one can write 
\begin{equation}
    \sum_{u \in \mathcal (H_0 \oplus f^{(a)})^\perp} (1-\frac{2\delta}{3})^{n-|u|}\left(\frac{\delta}{3} \right)^{|u|} =  \frac{1}{|\mathcal H_0 \oplus f^{(a)}|} \sum_{u \in \mathcal H_0 \oplus f^{(a)}}\left(1-\delta\right)^{|u|}.
\end{equation}

Thus the performance of triorthogonal qutrit codes can be characterized in terms of the classical weight enumerators of $\mathcal H_0$ and $\mathcal H_0 \oplus f^{(a)}$. Using these formulae one can reproduce the various results given in the main text.

\subsection{Our codes}
As an example, let us illustrate how the formulae from the previous subsection apply to our codes, for the case of maximum number of punctures $k=3m-2$. For this case, $H_0$ consists of two rows, and is of the form:
\begin{equation}
    H_0 = \begin{pmatrix}
    1 & 2 & 1 & 2 & \ldots & 1 & 2 & 0 & 1 & 2 & 0 & 1 & 2 \\
    0 & 0 & 0 & 0 & \ldots & 0 & 0 & 1 & 1 & 1 & 2 & 2 & 2
\end{pmatrix}.
\end{equation}
where $\begin{pmatrix} 1 & 2 \\ 0 & 0 \end{pmatrix}$ repeats $k$ times. 

The success probability is expressed in terms of the simple weight enumerator of the classical ternary code generated by $H_0$, and is easily seen to be,
\begin{equation}
    \begin{split}
    P_s  & = \frac{1}{9} \sum_{f \in \mathcal H_0}  \left(1-\delta \right)^{|u|} \\ & = \frac{1}{9}\left(1+ 2\left(1-\delta \right)^6 + 6\left(1-\delta \right)^{2k+4}\right), \\
    & \approx  1-\left(4+\frac{4 k}{3}\right)
   \delta +\frac{2}{3} \left(11+7
   k+2 k^2\right) \delta ^2 + O(\delta^3). \end{split}
\end{equation}

For any choice of $a$, $\mathcal H_0 \oplus f^{(a)}$ is generated by,
\begin{equation}
    H_0 \oplus f^{(a)} = \begin{pmatrix} 
    1 & 1 & 1 & 1 & \ldots & 1 & 1 & 1 & 1 & 1 & 1 & 1 & 1 \\
    1 & 2 & 1 & 2 & \ldots & 1 & 2 & 0 & 1 & 2 & 0 & 1 & 2 \\
    0 & 0 & 0 & 0 & \ldots & 0 & 0 & 1 & 1 & 1 & 2 & 2 & 2
\end{pmatrix}.
\end{equation}
The numerator of the fraction in Eq. \eqref{error-output-qutrit} is determined by the simple classical weight enumerator of the classical code determined by $H_0 \oplus f^{(a)}$ and is seen to be,
\begin{equation}
    \begin{split}
    &  \frac{1}{27} \sum_{f \in \mathcal H_0\oplus f^{a}}  \left(1-\delta \right)^{|u|} \\ & = \frac{1}{27}\left(1+ 2\left(1-\delta \right)^6 + 12\left(1-\delta \right)^{4+k} + 4(1-\delta)^{3 + 2 k}+ 6\left(1-\delta \right)^{2k+4} + 2 \left(1-\delta \right)^{6+2k} \right). \end{split}
\end{equation}
Putting these together, we find,
\begin{equation}
\begin{split}
    \bar{\delta}^{(a)} \approx  \left(1+\frac{k^2}{3}\right) \delta
   ^2+\left(2+2 k+\frac{k^2}{3}
   +\frac{k^3}{9}\right) \delta
   ^3+O\left(\delta ^4\right).
   \end{split}
\end{equation}
\bibliographystyle{quantum2}
\bibliography{qudit2023}

\end{document}